\documentclass[english]{article}
\pdfoutput=1
\usepackage[latin9]{inputenc}
\usepackage{graphicx}
\usepackage{amssymb}
\usepackage{esint}

\makeatletter

\usepackage{babel}
\makeatother

\begin{document}

\title{Quantum theory of optomechanical cooling}

\author{Florian Marquardt (1), A. A. Clerk (2), and S. M. Girvin (3)}

\maketitle
(1) Sektion Physik, Ludwig-Maximilians-Universität München, Theresienstr.
37, D-80333 München, Germany; (2) Department of Physics, McGill University,
3600 rue University, Montreal, QC Canada H3A 2T8; (3) Department of
Physics, Yale University, PO Box 208120, New Haven, CT 06520-8120

\begin{abstract}
We review the quantum theory of cooling of a mechanical oscillator
subject to the radiation pressure force due to light circulating inside
a driven optical cavity. Such optomechanical setups have been used
recently in a series of experiments by various groups to cool mechanical
oscillators (such as cantilevers) by factors reaching $10^{5}$, and
they may soon go to the ground state of mechanical motion. We emphasize
the importance of the sideband-resolved regime for ground state cooling,
where the cavity ring-down rate is smaller than the mechanical frequency.
Moreover, we illustrate the strong coupling regime, where the cooling
rate exceeds the cavity ring-down rate and where the driven cavity
resonance and the mechanical oscillation hybridize. Keywords: cavity
QED, optomechanics, micromechanics, sideband cooling, radiation pressure
\end{abstract}
\newcommand{\prl}{Phys. Rev. Lett.}

\newcommand{\pra}{Phys. Rev. A}

\section{Introduction}

The interaction of light with matter has been at the heart of the
development of quantum mechanics since its inception. As for the mechanical
effects of light, these become most pronounced in a setup where the
light intensity is resonantly enhanced (i.e. an optical cavity) and
where photons transfer maximum momentum to a mechanical object, e.g.
by being reflected multiple times from a movable mirror attached to
a cantilever. The study of radiation pressure effects on a movable
mirror was pioneered in seminal papers by Braginsky \cite{1967_BraginskyManukin_PonderomotiveEffectsEMRadiation,1970_Braginsky_OpticalCoolingExperiment}.
Strong changes of the mechanical properties of the mirror were observed
later in an experiment by the Walther group \cite{1983_10_DorselWalther_BistabilityMirror},
where a macroscopic mirror was found to exhibit two stable equilibrium
positions under the action of the cavity's radiation field. The most
recent series of activity in this field started with experiments observing
optomechanical cooling first using feedback \cite{1999_10_Cohadon_CoolingMirrorFeedback,2006_11_Bouwmeester_FeedbackCooling}
and later \cite{2003_08_Vogel_PhotothermalForceOnCantilever,2004_12_HoehbergerKarrai_CoolingMicroleverNature,2006_07_Arcizet_CoolingMirror,2006_05_AspelmeyerZeilinger_SelfCoolingMirror,2006_11_Kippenberg_RadPressureCooling,2006_12_NergisMavalvala_LIGO,2007_07_Harris_MembraneInTheMiddle}
using the intrinsic effect discussed in the following. In addition,
we note the self-induced optomechanical oscillations \cite{2005_02_MarquardtHarrisGirvin_Cavity}
that have been observed in radiation-pressure driven microtoroidal
optical resonators \cite{2005_06_Vahala_SelfOscillationsCavity,2005_07_VahalaTheoryPRL}
and other setups \cite{2004_KarraiConstanze_IEEE,2007_11_LudwigNeuenhahn_SelfInducedOscillations}.
For a recent review see \cite{2007_12_KippenbergVahala_ReviewOptomechanics}.
The study of these systems has been made even more fruitful by the
realization that the same (or essentially similar) physics may be
observed in systems ranging from driven LC circuits coupled to cantilevers
\cite{2007_Wineland_RFcircuitCooling} over superconducting single
electron transistors and microwave cavities coupled to nanobeams \cite{2005_11_Clerk_SSET_Cooling,2005_11_Blencowe_SET_NJP,2006_08_Schwab_CPB_Molasses,2006_BennettClerk_NEMSLaser,2007_Armour_ResonatorSSET,2007_Rodrigues_InstabilitySSET,2008_01_LehnertMicrowaveNanomechanics}
to clouds of cold atoms in an optical lattice, whose oscillations
couple to the light field \cite{2006_03_MeiserMeystre_CoupledDynamicsAtomsAndCantileverCavity,2007_08_Murch_AtomsCavityHeating}.
Cooling to the ground-state may open the door to various quantum effects
in these systems, including {}``cat'' states \cite{1999_05_Bose_Cat},
entanglement \cite{2003_09_Marshall_QSuperposMirror,2005_12_Pinard_EntanglingAndSelfCooling}
and Fock state detection \cite{2007_07_Harris_MembraneInTheMiddle}.
\begin{figure}
\begin{centering}
\includegraphics[width=0.9\columnwidth]{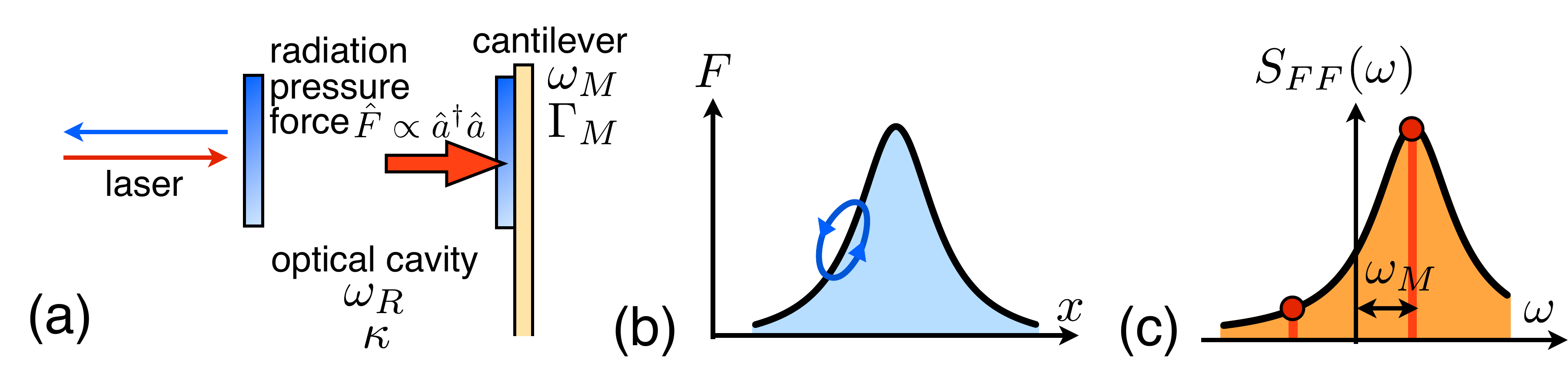}
\par\end{centering}

\caption{\label{setupfig}(a) The standard optomechanical setup treated in
the text: A driven optical cavity with a movable mirror. (b) Moving
the mirror in a cycle can result in work extracted by the light-field,
due to the finite cavity ring-down rate. (c) Radiation pressure force
noise spectrum.}

\end{figure}

All the intrinsic optomechanical cooling experiments are based on
the fact that the radiation field introduces extra damping for the
cantilever. In such a classical picture, the effective temperature
of the single mechanical mode of interest is related to the bath temperature
$T$ by $T_{{\rm eff}}/T=\Gamma_{M}/(\Gamma_{{\rm opt}}+\Gamma_{M})$,
where $\Gamma_{M}$ and $\Gamma_{{\rm opt}}$ are the intrinsic mechanical
damping rate and the optomechanical cooling rate, respectively. Thus
there is no limit to cooling in this regime, provided the laser power
(and thus the cooling rate $\Gamma_{{\rm opt}}$) can be increased
without any deleterious effects such as unwanted heating by absorption,
and provided the cooling rate remains sufficiently smaller than the
mechanical frequency and the cavity ring-down rate. However, at sufficiently
low temperatures, the unavoidable photon shot noise inside the cavity
counteracts cooling. To study the resulting quantum limits to cooling,
a fully quantum-mechanical theory is called for, which we provided
in Ref.~\cite{2007_01_Marquardt_CantileverCooling}, based on the
general quantum noise approach. Independently, a derivation emphasizing
the analogy to ion sideband cooling was developed in Ref.~\cite{2007_02_WilsonRae_Cooling}.
In the present paper, we will review and illustrate our theory. We
start by outlining the basic classical picture, then present the quantum
noise approach that provides a transparent and straightforward way
to derive cooling rates and quantum limits for the phonon number.
Finally, we illustrate the strong coupling regime that was first predicted
in our Ref.~\cite{2007_01_Marquardt_CantileverCooling}.

\section{Basic classical picture}

In this section, we briefly review the basic classical description
of optomechanical cooling.

The main ingredient for optomechanical cooling is the appearance of
extra damping, ideally without extra fluctuations. This damping is
introduced because the light-induced force reacts with a finite delay
time. In the case of radiation pressure, this comes about due to the
ring-down time $\kappa^{-1}$ of the cavity. On the other hand, bolometric
(i.e. photothermal) forces are produced when a bimorph cantilever
absorbs some of the radiation circulating inside the cavity. When
bolometric forces dominate, it is the finite time of thermal conductance
that sets the time-lag between the impinging radiation intensity and
the resulting change in the cantilever temperature, which is proportional
to the force. 

The physical picture behind damping is simplest when the time-lag
is small compared to the oscillation period of the cantilever. Then
both radiation pressure and bolometric forces give rise to the same
physics, modulo the appearance of a different time-scale in the two
cases. To describe this physics, we first fix our coordinate system:
Increasing the displacement $x$ of the cantilever means elongating
the cavity, and thus the optical resonance frequency (of the mode
of interest) decreases. Let us consider the cantilever being placed
at some location to the left of the resonance (Fig.~\ref{setupfig}).
This means that the optical mode frequency is still higher than the
frequency of the incoming laser radiation, which is therefore red-detuned
with respect to the optical resonance. Now imagine moving the cantilever
in a small cycle. As it moves towards the resonance with a finite
speed, the light-induced force does work on the cantilever. However,
due to the time-lag it remains smaller than it would be in the case
of infinitely slow (adiabatic) motion. Conversely, as the cantilever
moves back again in the second half of the cycle, the force extracts
energy and it is larger than for adiabatic motion. In total, the work
done during such a cycle by the light-induced force is negative, i.e.
mechanical energy is extracted from the cantilever (Fig.~\ref{setupfig}b).
This kind of physics may be modeled by writing down a simple relaxation-type
equation for the force, which tries to reach its proper $x$-dependent
value $\mathcal{F}(x)$ with some time-lag:

\begin{equation}
\dot{F}(t)=(\mathcal{F}(x(t))-F(t))/\tau.\label{eq:flag}\end{equation}
The cantilever is a damped harmonic oscillator driven by the light-induced
force:

\begin{equation}
m\ddot{x}=-m\omega_{M}^{2}(x-x_{0})-m\Gamma_{M}\dot{x}+F,\end{equation}
where $x_{0}$ is the mechanical equilibrium position, $m$ the cantilever
mass, $\omega_{M}$ its mechnical frequency, and $\Gamma_{M}$ the
intrinsic damping rate. Linearizing equation (\ref{eq:flag}) with
respect to small displacements from the mechanical equilibrium position
$\bar{x}$ and inserting it into the equation of motion of the cantilever
then yields the extra damping force. In Fourier space, where $x(t)=\int x[\omega]e^{-i\omega t}d\omega/2\pi$,
we find the following linearized equation of motion (at $\omega\neq0$):

\begin{equation}
-\omega^{2}mx[\omega]=-m\omega_{M}^{2}x[\omega]+im\omega\Gamma_{M}x[\omega]+\mathcal{F}'(\bar{x})x[\omega]/(1-i\omega\tau)\label{eq:linearizedsimple}\end{equation}
Comparing the last two terms (the intrinsic damping with the imaginary
part of the optomechanical term), we find that the optomechanical
damping rate is given by 

\begin{equation}
\Gamma_{{\rm opt}}=\frac{\mathcal{F}'(\bar{x})}{m\omega_{M}}\frac{\omega_{M}\tau}{1+(\omega_{M}\tau)^{2}},\end{equation}
for the simple ansatz of Eq.~(\ref{eq:flag}). According to this
analysis, one would expect the maximum effect to occur when $\omega_{M}\tau=1$,
i.e. when the time-delay matches the period of the cantilever motion.
As we will see further below, this conclusion is not upheld by the
full quantum-mechanical analysis for the case of radiation pressure.

Equation (\ref{eq:flag}) and the subsequent analysis holds exactly
for the bolometric force. In that case, $\tau$ is the finite time
of thermal conductance and $\mathcal{F}(x)=\mathcal{F}_{{\rm max}}I(x)/I_{{\rm max}}$
is the displacement-dependent bolometric force, where $I(x)=I_{{\rm max}}/(1+(2\Delta(x)/\kappa)^{2})$
is the intensity profile and $\Delta(x)=x\omega_{R}/L$ is the position-dependent
detuning between incoming laser radiation and optical resonance at
$x=0$. For the radiation-pressure force, we can use the present analysis
only in the regime of $\kappa\gg\omega_{M}$, and only if we allow
for a position dependent relaxation rate $1/\tau$ (see \cite{2007_01_Marquardt_CantileverCooling}).

In both cases, however, the shape of $\Gamma_{{\rm opt}}$ as a function
of cantilever position $\bar{x}$ is determined by the slope of the
intensity profile, i.e. in particular by the sign of $\mathcal{F}'$.
To the left of the resonance, where $\mathcal{F}'>0$, we indeed obtain
extra damping: $\Gamma_{{\rm opt}}>0$. As long as there are no extra
fluctuations introduced by the light-induced force (i.e. if we may
disregard shot noise), the effective temperature of the mechanical
degree of freedom is therefore reduced according to the ratio of intrinsic
and optomechanical damping rates:

\begin{equation}
T_{{\rm eff}}=T\frac{\Gamma_{M}}{\Gamma_{M}+\Gamma_{{\rm opt}}}.\label{eq:effT}\end{equation}
This can be obtained, for example, by solving the Langevin equation
that includes the thermal fluctuations of the mechanical heat bath
(whose strength is set by $\Gamma_{M}$ according to the fluctuation-dissipation
theorem). Then the effective temperature may be defined according
to the equipartition theorem: $m\tilde{\omega}_{M}^{2}\left\langle (x-\bar{x})^{2}\right\rangle =k_{B}T_{{\rm eff}}$,
where $\tilde{\omega}_{M}$ contains the frequency renormalization
due to the real part of the optomechanical term in Eq.~(\ref{eq:linearizedsimple}).

\section{Quantum noise approach}

In the quantum regime, we have to take into account the shot noise
that tends to heat the cantilever and enforces a finite quantum limit
for the cantilever phonon number. The quantum picture can also be
understood as Raman scattering: Incoming photons, red-detuned with
respect to the optical resonance, absorb a phonon from the cantilever,
thereby cooling it. However, there is also a finite probability for
phonon emission, and thus heating. The purpose of a quantum theory
is to discuss the balance of these effects.

The idea behind the quantum noise approach to quantum-dissipative
systems is to describe the environment fully by the correlator of
the fluctuating force that couples to the quantum system of interest.
If the coupling is weak enough, knowledge of the correlator is sufficient
to fully describe the influence of the environment. In our case, this
means looking at the spectrum of the radiation pressure force fluctuations,
which are produced by the shot noise of photons inside the driven
optical cavity mode, i.e. a nonequilibrium environment. In other applications,
we might be dealing with the electrical field fluctuations produced,
e.g., by a driven electronic circuit (superconducting single electron
transistor, quantum point contact, LC circuit) capacitively coupled
to some nanobeam. The general formulas remain the same for all of
these cases, and only the noise spectrum changes.

The Fourier transform of the force correlator defines the spectral
noise density:

\begin{equation}
S_{FF}(\omega)=\int dt\, e^{i\omega t}\left\langle \hat{F}(t)\hat{F}(0)\right\rangle \,.\end{equation}
The noise spectrum $S_{FF}$ is real-valued and non-negative. However,
in contrast to the classical case, it is asymmetric in frequency,
since $\hat{F}(t)$ and $\hat{F}(0)$ do not commute. This asymmetry
has an important physical meaning: Contributions at positive frequencies
indicate the possibility of the environment to absorb energy, while
those at negative frequencies imply its ability to release energy
(to the cantilever). All the optomechanical effects can be described
in terms of $S_{FF}$, as long as the coupling is weak.

\begin{figure}
\begin{centering}
\includegraphics[width=1\columnwidth]{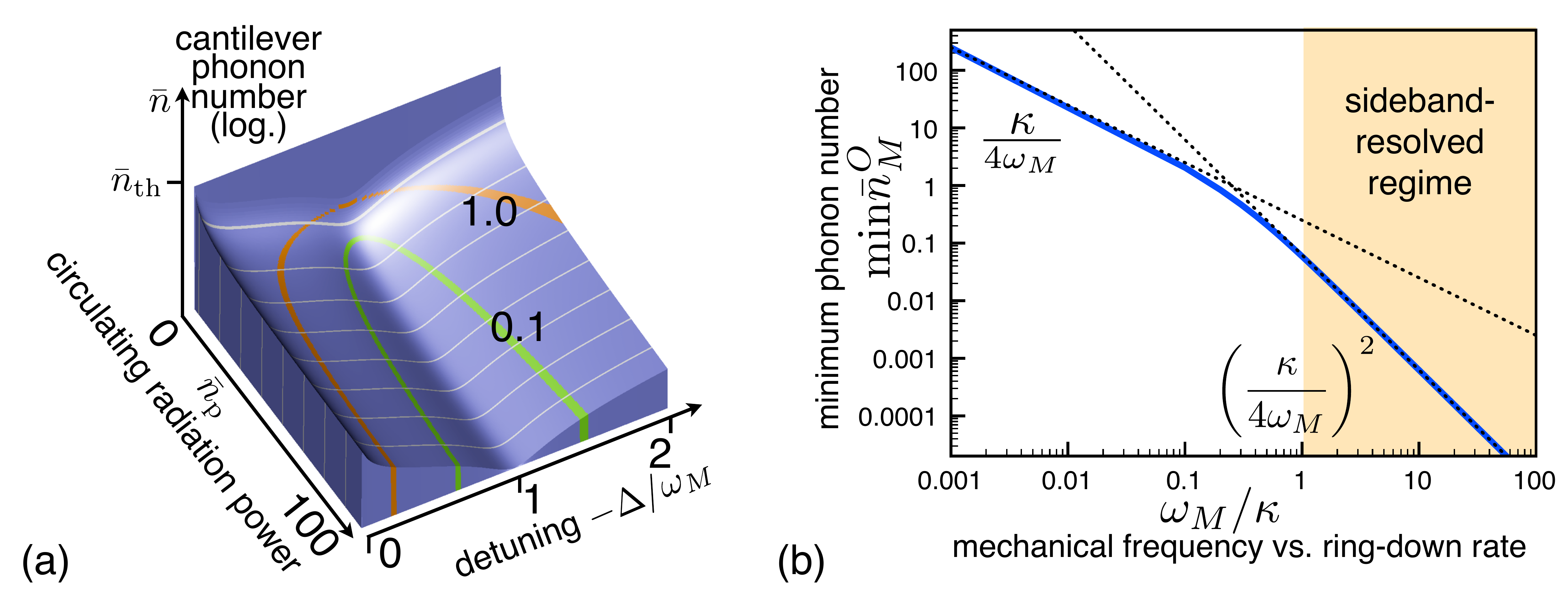}
\par\end{centering}

\caption{\label{cool3D}(a) Cantilever phonon number as a function of circulating
radiation power inside the cavity and as a function of detuning between
laser and optical resonance. The mean phonon number in steady state
is plotted on a logarithmic scale. Contour lines indicate $\bar{n}=1$
and $\bar{n}=0.1$. For this plot, the following parameters have been
used: $\omega_{M}/\kappa=0.3,\,\bar{n}_{{\rm th}}=10^{3},\, Q=\omega_{M}/\Gamma_{M}=10^{6}$,
and $(\omega_{R}/\omega_{M})(x_{{\rm ZPF}}/L)\approx0.012$. (b) The
minimum phonon number as a function of the ratio between the mechanical
frequency $\omega_{M}$ and the optical cavity's ring-down rate $\kappa$,
according to Eq.~(\ref{eq:MinimalPhon}). Ground-state cooling is
possible in the regime $\omega_{M}\gg\kappa$, i.e. the {}``good
cavity'' or {}``resolved sideband'' limit.}

\end{figure}
The optomechanical damping rate is given by the difference of noise
spectra at positive and negative frequencies,

\begin{equation}
\Gamma_{{\rm opt}}=\frac{x_{{\rm ZPF}}^{2}}{\hbar^{2}}[S_{FF}(\omega_{M})-S_{FF}(-\omega_{M})]\,.\label{eq:Gopt}\end{equation}
This formula is obtained by applying Fermi's Golden Rule to derive
the transition rates arising from the coupling of the cantilever to
the light field, i.e. from the term $\hat{H}_{{\rm int}}=-\hat{F}\hat{x}$
in the Hamiltonian. These are

\begin{equation}
\Gamma_{\downarrow}^{{\rm opt}}=\frac{x_{{\rm ZPF}}^{2}}{\hbar^{2}}S_{FF}(\omega_{M})\,\,\,\,\,,\,\,\,\,\,\,\Gamma_{\uparrow}^{{\rm opt}}=\frac{x_{{\rm ZPF}}^{2}}{\hbar^{2}}S_{FF}(-\omega_{M}).\end{equation}
These rates enter the complete master equation for the density matrix
$\hat{\rho}$ of the cantilever in the presence of the equilibrium
heat bath (that would lead to a thermal population $\bar{n}_{{\rm th}}$)
and the radiation field:

\begin{equation}
\dot{\hat{\rho}}=\left[(\Gamma_{\downarrow}^{{\rm opt}}+\Gamma_{M}(\bar{n}_{{\rm th}}+1))\mathcal{D}[\hat{a}]+(\Gamma_{\uparrow}^{{\rm opt}}+\Gamma_{M}\bar{n}_{{\rm th}})\mathcal{D}[\hat{a}^{\dagger}]\right]\hat{\rho}\end{equation}
Here the equation has been written in the interaction picture (disregarding
the oscillations at $\tilde{\omega}_{M}$), and

\begin{equation}
\mathcal{D}[\hat{A}]\hat{\rho}=\frac{1}{2}(2\hat{A}\hat{\rho}\hat{A}^{\dagger}-\hat{A}^{\dagger}\hat{A}\hat{\rho}-\hat{\rho}\hat{A}^{\dagger}\hat{A})\end{equation}
is the standard Lindblad operator for downward ($\hat{A}=\hat{a}$)
or upward ($\hat{A}=\hat{a}^{\dagger}$) transitions in the oscillator.
Restricting ourselves to the populations $\rho_{nn}$, we obtain the
equation for the phonon number $\bar{n}=\left\langle \hat{n}\right\rangle =\sum_{n}n\rho_{nn}$:

\begin{equation}
\dot{\bar{n}}=\Gamma_{M}\bar{n}_{{\rm th}}+\Gamma_{\uparrow}^{{\rm opt}}-(\Gamma_{M}+\Gamma_{{\rm opt}})\bar{n},\end{equation}
which yields the steady-state phonon number in the presence of optomechanical
cooling:

\begin{equation}
\bar{n}_{M}=\frac{\Gamma_{M}\bar{n}_{{\rm th}}+\Gamma_{{\rm opt}}\bar{n}_{M}^{O}}{\Gamma_{M}+\Gamma_{{\rm opt}}}.\end{equation}
This is the weighted average of the thermal and the optomechanical
phonon numbers. It represents the correct generalization of the classical
formula for the effective temperature, Eq.~(\ref{eq:effT}). Here

\begin{equation}
\bar{n}_{M}^{O}=\frac{\Gamma_{\uparrow}^{{\rm opt}}}{\Gamma_{{\rm opt}}}=\frac{1}{\Gamma_{\downarrow}^{{\rm opt}}/\Gamma_{\uparrow}^{{\rm opt}}-1}=\left[\frac{S_{FF}(\omega_{M})}{S_{FF}(-\omega_{M})}-1\right]^{-1}\label{eq:nMO}\end{equation}
is the minimal phonon number reachable by optomechanical cooling.
This quantum limit is reached when $\Gamma_{{\rm opt}}\gg\Gamma_{M}$.
Then, the cooling effect due to extra damping is balanced by the shot
noise in the cavity, which leads to heating.

The radiation pressure force is proportional to the photon number:
$\hat{F}=(\hbar\omega_{R}/L)\hat{a}^{\dagger}\hat{a}$. A brief calculation
for the photon number correlator inside a driven cavity \cite{2007_01_Marquardt_CantileverCooling}
yields its spectrum in the form of a Lorentzian that is shifted by
the detuning $\Delta=\omega_{L}-\omega_{R}$ between laser and optical
resonance frequency $\omega_{R}$:

\begin{equation}
S_{FF}(\omega)=\left(\frac{\hbar\omega_{R}}{L}\right)^{2}\bar{n}_{{\rm p}}\frac{\kappa}{(\omega+\Delta)^{2}+(\kappa/2)^{2}}\,,\end{equation}
where $\bar{n}_{p}$ is the photon number circulating inside the cavity.
A plot of the resulting steady-state phonon number is shown in Fig.~\ref{cool3D}a.

Inserting this spectrum into Eqs.~(\ref{eq:Gopt}) and (\ref{eq:nMO})
yields the optomechanical cooling rate and the minimum phonon number
as a function of detuning $\Delta$. The minimum of $\bar{n}_{M}^{O}$
is reached at a detuning $\Delta=-\sqrt{\omega_{M}^{2}+(\kappa/2)^{2}}$,
and it is (see Fig.~\ref{cool3D}b):

\begin{equation}
{\rm min}\,\bar{n}_{M}^{O}=\frac{1}{2}\left(\sqrt{1+\left(\frac{\kappa}{2\omega_{M}}\right)^{2}}-1\right)\,.\label{eq:MinimalPhon}\end{equation}
For slow cantilevers, $\omega_{M}\ll\kappa$, we have ${\rm min}\,\bar{n}_{M}^{O}=\kappa/(4\omega_{M})\gg1$.
Ground-state cooling becomes possible for high-frequency cantilevers
(and/or high-finesse cavities), when $\kappa\ll\omega_{M}$. Then,
we find

\begin{equation}
\min\,\bar{n}_{M}^{O}\approx\left(\frac{\kappa}{4\omega_{M}}\right)^{2}\,.\end{equation}
As explained in Ref.~\cite{2007_02_WilsonRae_Cooling}, these two
regimes can be brought directly into correspondence with the known
regimes for laser-cooling of harmonically bound atoms, namely the
Doppler limit for $\omega_{M}\ll\kappa$ and the resolved sideband
regime for $\omega_{M}\gg\kappa$.

\section{Strong coupling effects}

Up to now, we have assumed that the coupling between light and mechanical
degree of freedom is sufficiently weak to allow for a solution in
terms of a master equation, employing the rates obtained from the
quantum noise approach. However, as the coupling becomes stronger
(e.g. by increasing the laser input power), $\Gamma_{{\rm opt}}$
may reach the cavity decay rate $\kappa$. Then, the spectrum of force
fluctuations is itself modified by the presence of the cantilever.
It becomes necessary to solve for the coupled dynamics of the light
field and the mechanical motion. This has been done in Ref.~\cite{2007_01_Marquardt_CantileverCooling},
by writing down the Heisenberg equations of motion for the cantilever
and the optical mode, and solving them after linearization. Here we
will only discuss the result.

To analyze these features, let us consider the spectrum of the cantilever
motion,

\begin{equation}
S_{cc}(\omega)=\int dt\, e^{i\omega t}\left\langle \hat{c}^{\dagger}(t)\hat{c}(0)\right\rangle \,,\end{equation}
where $\hat{c}$ is the annihilation operator for the cantilever harmonic
oscillator. At weak coupling, this spectrum displays a peak at the
(renormalized) cantilever frequency, i.e. at $\omega=-\tilde{\omega}_{M}$
{[}the minus sign is a consequence of our choice of the definition,
following \cite{2007_01_Marquardt_CantileverCooling}]. Its width
(FWHM) is given by $\Gamma_{{\rm opt}}+\Gamma_{M}$, and its total
weight $\int S_{cc}(\omega)d\omega/2\pi$ yields the phonon number
$\bar{n}=\left\langle \hat{c}^{\dagger}\hat{c}\right\rangle $. As
the laser power is increased, the width increases and the weight diminishes
(in the cooling regime). %
\begin{figure}
\begin{centering}
\includegraphics[width=0.6\columnwidth]{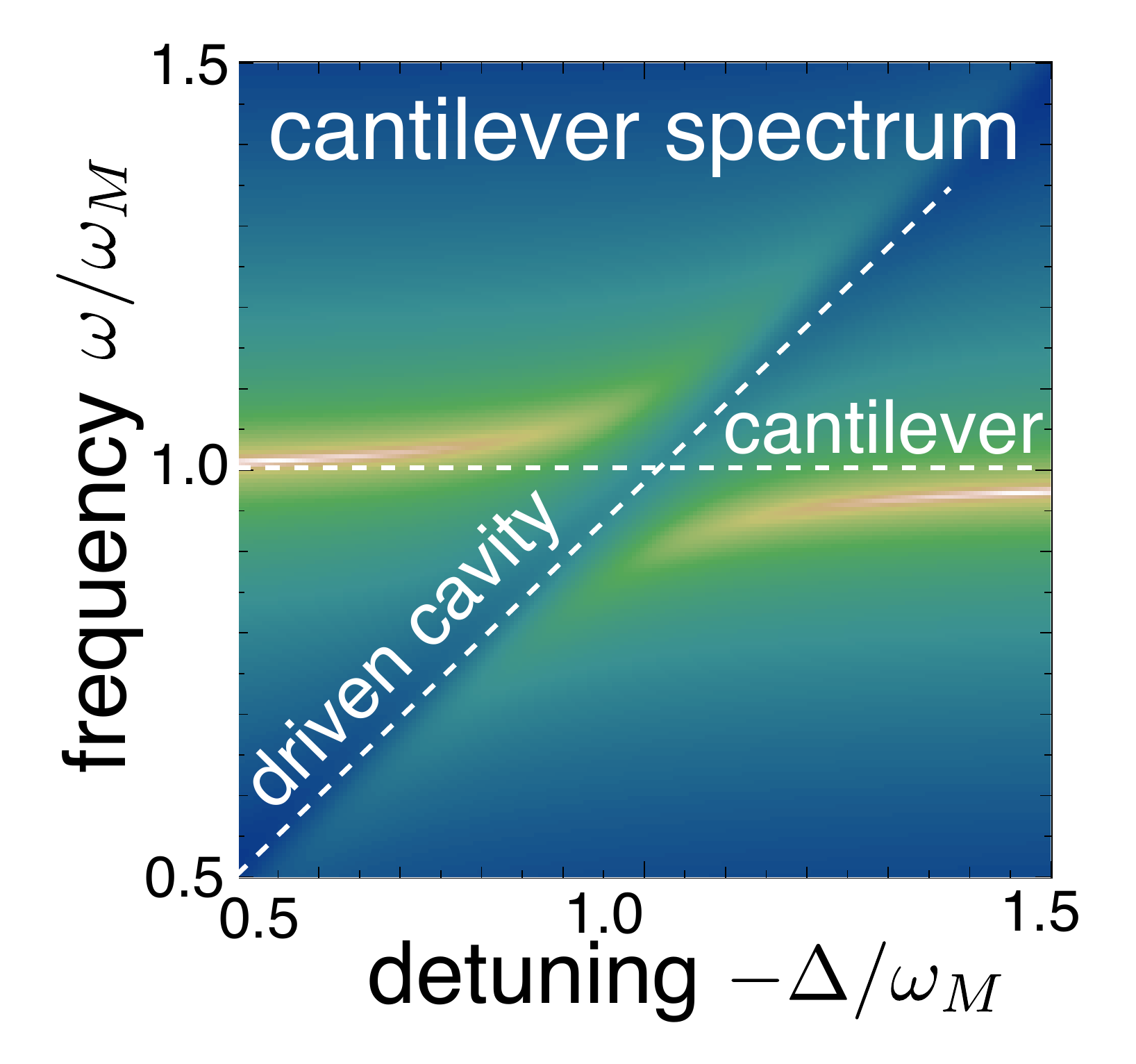}
\par\end{centering}

\caption{\label{fig3}The spectrum $S_{cc}(-\omega)$ of the cantilever motion,
as a function of detuning $\Delta$ and spectral frequency $\omega$,
for fixed circulating power. For the plot shown here, the strong-coupling
regime is reached, i.e. $\Gamma_{{\rm opt}}/\kappa$ becomes larger
than one near resonance. Consequently, the driven cavity resonance
and the cantilever resonance hybridize, and an avoided crossing is
observed. The parameters for this plot are: $\kappa/\omega_{M}=0.1$,
$\Gamma_{M}/\omega_{M}=10^{-5}$, $\left(\omega_{R}/\omega_{M}\right)\left(x_{{\rm ZPF}}/L\right)=0.01$,
$\bar{n}_{p}=100$, and $\bar{n}_{{\rm th}}=10^{3}$.}

\end{figure}

When $\Gamma_{{\rm opt}}/\kappa$ is no longer much smaller than one,
deviations from the weak-coupling results start to appear \cite{2007_01_Marquardt_CantileverCooling}.
The most dramatic effect is observed when $\Gamma_{{\rm opt}}/\kappa>1/2$:
The peak splits into two (Fig.~\ref{fig3}). As it were, in this
strong-coupling regime one actually encounters the hybridization of
two coupled harmonic oscillators, namely the cantilever and the driven
optical mode (with an effective frequency set by the detuning). At
resonance, i.e. for $\Delta=-\omega_{M}$, the splitting is set by
$2\alpha$, where the coupling frequency $\alpha$ is determined by
the circulating laser power, the ratio of mechanical zero-point fluctuations
to the cavity length, and by the optical resonance frequency: 

\begin{equation}
\alpha=\omega_{R}\sqrt{\bar{n}_{p}}\frac{x_{{\rm ZPF}}}{L}\,.\end{equation}

\section{Outlook}

During the past year, several new ideas have been introduced into
the field of optomechanical cooling. For example, placing a movable
membrane in the middle of a standard optical cavity \cite{2007_07_Harris_MembraneInTheMiddle}
can lead to orders of magnitude better performance, as it separates
the mechanical from the optical elements. Such a setup has been used
to cool from $300K$ down to $7mK$, and it may ultimately by employed
for Fock state detection of mechanical vibrations \cite{2007_07_Harris_MembraneInTheMiddle}.
Even nanomechanical objects (such as nanowires) might be placed inside
the standing light wave \cite{2007_07_FaveroKarrai_ScatteringCooling}
and cooled by scattering. Doppler cooling of Bragg mirrors may provide
another promising approach \cite{2007_06_Karrai_DopplerCooling}.
Furthermore, sideband-resolved cooling with $\omega_{M}/\kappa\sim20$
has been demonstrated recently \cite{2007_09_Kippenberg_SidebandCooling},
paving the way for ground-state cooling when combined with cryogenics
\cite{2008_03_Aspelmeyer_CryoCooling}.

\section{Acknowledgements}

We thank Jack Harris for sharing his many insights about optomechanical
systems. F.~M. also thanks B.~Kubala, M.~Ludwig and C.~Neuenhahn
for discussions on this topic. F.~M. thanks the DFG for support via
SFB 631, the Emmy-Noether program, and the NIM cluster of excellence.
The work of S.~M.~G. was supported by the NSF via grants DMR-0603369
and DMR-0653377. A.~A.~C. would like to thank the Canadian Institute
for Advanced Research (CIFAR).

\bibliographystyle{unsrt}
\bibliography{/Users/florian/pre/bib/BibFM}

\begin{thebibliography}{10}

\bibitem{1967_BraginskyManukin_PonderomotiveEffectsEMRadiation}
V.B. Braginsky and A.B. Manukin.
\newblock Ponderomotive effects of electromagnetic radiation.
\newblock {\em Soviet Physics JETP}, 25:653, 1967.

\bibitem{1970_Braginsky_OpticalCoolingExperiment}
V.~B. Braginsky, A.~B. Manukin, and M.~Yu. Tikhonov.
\newblock Investigation of dissipative ponderomotive effects of electromagnetic
  radiation.
\newblock {\em Soviet Physics JETP}, 31:829, 1970.

\bibitem{1983_10_DorselWalther_BistabilityMirror}
A.~Dorsel, J.~D. McCullen, P.~Meystre, E.~Vignes, and H.~Walther.
\newblock Optical bistability and mirror confinement induced by radiation
  pressure.
\newblock {\em \prl}, 51:1550, 1983.

\bibitem{1999_10_Cohadon_CoolingMirrorFeedback}
P.~F. Cohadon, A.~Heidmann, and M.~Pinard.
\newblock Cooling of a mirror by radiation pressure.
\newblock {\em \prl}, 83:3174, 1999.

\bibitem{2006_11_Bouwmeester_FeedbackCooling}
D.~Kleckner and D.~Bouwmeester.
\newblock Sub-kelvin optical cooling of a micromechanical resonator.
\newblock {\em Nature}, 444:75, 2006.

\bibitem{2003_08_Vogel_PhotothermalForceOnCantilever}
M.~Vogel, C.~Mooser, K.~Karrai, and R.~J. Warburton.
\newblock Optically tunable mechanics of microlevers.
\newblock {\em Applied Physics Letters}, 83:1337, 2003.

\bibitem{2004_12_HoehbergerKarrai_CoolingMicroleverNature}
C.~{H\"ohberger}-Metzger and K.~Karrai.
\newblock Cavity cooling of a microlever.
\newblock {\em Nature}, 432:1002, 2004.

\bibitem{2006_07_Arcizet_CoolingMirror}
O.~Arcizet, P.~F. Cohadon, T.~Briant, M.~Pinard, and A.~Heidmann.
\newblock Radiation-pressure cooling and optomechanical instability of a
  micro-mirror.
\newblock {\em Nature}, 444:71, 2006.

\bibitem{2006_05_AspelmeyerZeilinger_SelfCoolingMirror}
S.~Gigan et~al.
\newblock Self-cooling of a micromirror by radiation pressure.
\newblock {\em Nature}, 444:67, 2006.

\bibitem{2006_11_Kippenberg_RadPressureCooling}
A.~Schliesser, P.~Del'Haye, N.~Nooshi, K.~J. Vahala, and T.~J. Kippenberg.
\newblock Cooling of a micro-mechanical oscillator using radiation pressure
  induced dynamical back-action.
\newblock {\em \prl}, 97:243905, 2006.

\bibitem{2006_12_NergisMavalvala_LIGO}
Thomas Corbitt et~al.
\newblock Toward achieving the quantum ground state of a gram-scale mirror
  oscillator.
\newblock {\em \prl}, 98:150802, 2007.

\bibitem{2007_07_Harris_MembraneInTheMiddle}
J.~D. Thompson, B.~M. Zwickl, A.~M. Jayich, F.~Marquardt, S.~M. Girvin, and
  J.~G.~E. Harris.
\newblock Strong dispersive coupling of a high finesse cavity to a
  michromechanical membrane.
\newblock {\em Nature}, 452:72, 2008.

\bibitem{2005_02_MarquardtHarrisGirvin_Cavity}
F.~Marquardt, J.~G.~E. Harris, and S.~M. Girvin.
\newblock Dynamical multistability induced by radiation pressure in
  high-finesse micromechanical optical cavities.
\newblock {\em \prl}, 96:103901, 2006.

\bibitem{2005_06_Vahala_SelfOscillationsCavity}
T.~Carmon, H.~Rokhsari, L.~Yang, T.~J. Kippenberg, and K.~J. Vahala.
\newblock Temporal behavior of radiation-pressure-induced vibrations of an
  optical microcavity phonon mode.
\newblock {\em \prl}, 94:223902, 2005.

\bibitem{2005_07_VahalaTheoryPRL}
T.~J. Kippenberg, H.~Rokhsari, T.~Carmon, A.~Scherer, and K.~J. Vahala.
\newblock Analysis of radiation-pressure induced mechanical oscillation of an
  optical microcavity.
\newblock {\em \prl}, 95:033901, 2005.

\bibitem{2004_KarraiConstanze_IEEE}
C.~{H\"ohberger} and K.~Karrai.
\newblock Self-oscillation of micromechanical resonators.
\newblock {\em Nanotechnology 2004, Proceedings of the 4th IEEE conference on
  nanotechnology}, page 419, 2004.

\bibitem{2007_11_LudwigNeuenhahn_SelfInducedOscillations}
Max Ludwig, Clemens Neuenhahn, Constanze Metzger, Alexander Ortlieb, Ivan
  Favero, Khaled Karrai, and Florian Marquardt.
\newblock Self-induced oscillations in an optomechanical system.
\newblock {\em arXiv:0711.2661}, 2007.

\bibitem{2007_12_KippenbergVahala_ReviewOptomechanics}
T.~J. Kippenberg and K.~J. Vahala.
\newblock Cavity opto-mechanics.
\newblock {\em Optics Express}, 15:17172, 2007.

\bibitem{2007_Wineland_RFcircuitCooling}
K.~R. Brown, J.~Britton, R.~J. Epstein, J.~Chiaverini, D.~Leibfried, and D.~J.
  Wineland.
\newblock Passive cooling of a micromechanical oscillator with a resonant
  electric circuit.
\newblock {\em \prl}, 99:137205, 2007.

\bibitem{2005_11_Clerk_SSET_Cooling}
A.~A. Clerk and S.~Bennett.
\newblock Quantum nanoelectromechanics with electrons, quasi-particles and
  {Cooper} pairs: effective bath descriptions and strong feedback effects.
\newblock {\em New Journal of Physics}, 7:238, 2005.

\bibitem{2005_11_Blencowe_SET_NJP}
M.~P. Blencowe, J.~Imbers, and A.~D. Armour.
\newblock Dynamics of a nanomechanical resonator coupled to a superconducting
  single-electron transistor.
\newblock {\em New Journal of Physics}, 7:236, 2005.

\bibitem{2006_08_Schwab_CPB_Molasses}
A.~Naik et~al.
\newblock Cooling a nanomechanical resonator with quantum back-action.
\newblock {\em Nature}, 443:193, 2006.

\bibitem{2006_BennettClerk_NEMSLaser}
S.~D. Bennett and A.~A. Clerk.
\newblock Laser-like instabilities in quantum nano-electromechanical systems.
\newblock {\em Phys. Rev. B}, 74:201301, 2006.

\bibitem{2007_Armour_ResonatorSSET}
D.~A. Rodrigues, J.~Imbers, and A.~D. Armour.
\newblock Quantum dynamics of a resonator driven by a superconducting
  single-electron transistor: a solid-state analogue of the micromaser.
\newblock {\em \prl}, 98:067204, 2007.

\bibitem{2007_Rodrigues_InstabilitySSET}
D.~A. Rodrigues, J.~Imbers, T.~J. Harvey, and A.~D. Armour.
\newblock Dynamical instabilities of a resonator driven by a superconducting
  single-electron transistor.
\newblock {\em New Journal of Physics}, 9:84, 2007.

\bibitem{2008_01_LehnertMicrowaveNanomechanics}
C.~A. Regal, J.~D. Teufel, and K.~W. Lehnert.
\newblock Measuring nanomechanical motion with a microwave cavity
  interferometer.
\newblock {\em arXiv:0801.1827}, 2008.

\bibitem{2006_03_MeiserMeystre_CoupledDynamicsAtomsAndCantileverCavity}
D.~Meiser and P.~Meystre.
\newblock Coupled dynamics of atoms and radiation-pressure-driven
  interferometers.
\newblock {\em \pra}, 73:033417, 2006.

\bibitem{2007_08_Murch_AtomsCavityHeating}
K.~W. Murch, K.~L. Moore, S.~Gupta, and D.~M. Stamper-Kurn.
\newblock {Measurement of Intracavity Quantum Fluctuations of Light Using an
  Atomic Fluctuation Bolometer}.
\newblock {\em arXiv:0706.1005v2}, 2007.

\bibitem{1999_05_Bose_Cat}
S.~Bose, K.~Jacobs, and P.~L. Knight.
\newblock Scheme to probe the decoherence of a macroscopic object.
\newblock {\em \pra}, 59:3204, 1999.

\bibitem{2003_09_Marshall_QSuperposMirror}
William Marshall, Christoph Simon, Roger Penrose, and Dik Bouwmeester.
\newblock Towards quantum superpositions of a mirror.
\newblock {\em Physical Review Letters}, 91(13):130401, 2003.

\bibitem{2005_12_Pinard_EntanglingAndSelfCooling}
M.~Pinard, A.~Dantan, D.~Vitali, O.~Arcizet, T.~Briant, and A.~Heidmann.
\newblock Entangling movable mirrors in a double-cavity system.
\newblock {\em Europhysics Letters}, 72:747, 2005.

\bibitem{2007_01_Marquardt_CantileverCooling}
F.~Marquardt, J.~P. Chen, A.~A. Clerk, and S.~M. Girvin.
\newblock Quantum theory of cavity-assisted sideband cooling of mechanical
  motion.
\newblock {\em \prl}, 99:093902, 2007.

\bibitem{2007_02_WilsonRae_Cooling}
I.~Wilson-Rae, N.~Nooshi, W.~Zwerger, and T.~J. Kippenberg.
\newblock Theory of ground state cooling of a mechanical oscillator using
  dynamical back-action.
\newblock {\em \prl}, 99:093901, 2007.

\bibitem{2007_07_FaveroKarrai_ScatteringCooling}
I.~Favero and K.~Karrai.
\newblock Cavity cooling of a nanomechanical resonator by light scattering.
\newblock {\em arXiv:0707.3117}, 2007.

\bibitem{2007_06_Karrai_DopplerCooling}
K.~Karrai, I.~Favero, and C.~Metzger.
\newblock Doppler controlled dynamics of a mirror attached to a spring.
\newblock {\em arXiv:0706.2841}, 2007.

\bibitem{2007_09_Kippenberg_SidebandCooling}
A.~Schliesser, R.~Riviere, G.~Anetsberger, O.~Arcizet, and T.~J. Kippenberg.
\newblock Resolved sideband cooling of a micromechanical oscillator.
\newblock {\em arXiv:0709.4036}, 2007.

\bibitem{2008_03_Aspelmeyer_CryoCooling}
S.~{Gr\"oblacher}, S.~Gigan, H.~R. {B\"ohm}, A.~Zeilinger, and M.~Aspelmeyer.
\newblock Radiation-pressure self-cooling of a micromirror in a cryogenic
  environment.
\newblock {\em Europhys. Lett.}, 81:54003, 2008.

\end{thebibliography}

\end{document}